\documentclass[twocolumn]{article}

\usepackage{graphicx}
\usepackage{amsmath}
\usepackage{amssymb}
\usepackage{bigstrut}
\usepackage{tabularx,ragged2e,booktabs,caption,authblk}
\newcolumntype{C}[1]{>{\Centering}m{#1}}

\everymath{\displaystyle}
\usepackage{fullpage}
\usepackage{cite}
\usepackage{float}

\title{Anticipation induces polarized collective motion in attraction based models}

\author[1,2]{Daniel Str\"{o}mbom\thanks{\mbox{Corresponding author: daniel.strombom@math.uu.se}}}
\author[3]{Alice Antia}
\affil[1]{Department of Mathematics, Uppsala University, Uppsala, Sweden}
\affil[2]{Department of Biosciences, Swansea University, Swansea, UK}
\affil[3]{Department of Mathematics and Statistics, Carleton College, MN, USA}

\date{}
\begin{document}
\maketitle
\begin{abstract} 
{In most models of collective motion in animal groups each individual updates its heading based on the current positions and headings of its neighbors. Several authors have investigated the effects of including anticipation into models of this type, and have found that anticipation inhibits polarized collective motion in alignment based models and promotes milling and swarming in the one attraction-repulsion model studied. However, it was recently reported that polarized collective motion does emerge in an alignment based asynchronous lattice model with mutual anticipation. To our knowledge this is the only reported case where polarized collective motion has been observed in a model with anticipation. Here we show that including anticipation induces polarized collective motion in a synchronous, off lattice, attraction based model. This establishes that neither asynchrony, mutual anticipation nor motion restricted to a lattice environment are strict requirements for anticipation to promote polarized collective motion. In addition, unlike alignment based models the attraction based model used here does not produce any type of polarized collective motion in the absence of anticipation. Here anticipation is a direct polarization inducing mechanism. We believe that utilizing anticipation instead of frequently used alternatives such as explicit alignment terms, asynchronous updates and asymmetric interactions to generate polarized collective motion may be advantageous in some cases.
} 
\end{abstract}

\section*{Introduction} Self-propelled particle (SPP) models have been used to model collective motion in a large variety of animal groups including fish schools, bird flocks, herds of sheep and human crowds \cite{Vicsek2012}. In most SPP models each particle calculates its new heading and position based on the current positions and/or headings of its neighbors. Recently a number of studies have highlighted the potential importance of anticipation in models of this type \cite{Morin2015,Gerlee2016,Baggaley2016,Murakami2017}. In models including anticipation each particle uses the future anticipated positions and/or headings of its neighbors to calculate its new heading, rather than their current positions and headings. The effects of anticipation on group formation in alignment based Vicsek type models \cite{Vicsek1995} has been described in \cite{Morin2015,Baggaley2016} and the effects on an attraction-repulsion model  \cite{DOrsogna2006} has been described in \cite{Gerlee2016}. Interestingly, in both types of models it was established that including anticipation inhibits polarized collective motion and promote, or stabilize, milling and swarming. This is particularly surprising in the case of alignment based models because the production of polarized collective motion is a key feature of these models, whereas production of mills and swarms is not. In contrast it was recently shown that mutual anticipation allowed for polarized collective motion to emerge in an alignment based asynchronous lattice model \cite{Murakami2017}.

How general are these findings? In particular, does anticipation tend to promote swarming and milling in SPP models unless specific additional implementation choices are made? For example, is asynchrony, motion restricted to a lattice or mutual anticipation necessary requirements for polarized collective motion to emerge in models with anticipation? To address this question we suggest studying a family of models known to produce all three group types involved: mills, swarms and polarized groups. The local attraction model (LAM), where particles interact via local attraction alone, is the simplest model known to produce all three of these groups \cite{Strombom2011}. In addition, polarized group formation can be switched off in the LAM by implementing a fully synchronous version of the model. In this case only no group, mills and swarms are produced \cite{Strombom2017}. This latter version is a synchronous, off lattice, attraction only model that does not normally produce polarized collective motion. If polarized groups emerge when neighborhood wide anticipation is incorporated into this version of the LAM we will have established that neither asynchrony, restriction to on lattice motion, or mutual anticipation are necessary requirements for polarized collective motion to emerge in a model with anticipation. Complementing the findings in \cite{Murakami2017}. In addition, as mills and swarms are readily produced in well defined parameter regions of this model without anticipation we may also be able to determine whether anticipation stabilizes, promotes, or inhibits milling and/or swarming in this setting. Complementing the results in \cite{Morin2015,Gerlee2016,Baggaley2016}.

\section*{Model and Methods}
\begin{figure}[h!]
\centering
\includegraphics[width=0.65\columnwidth]{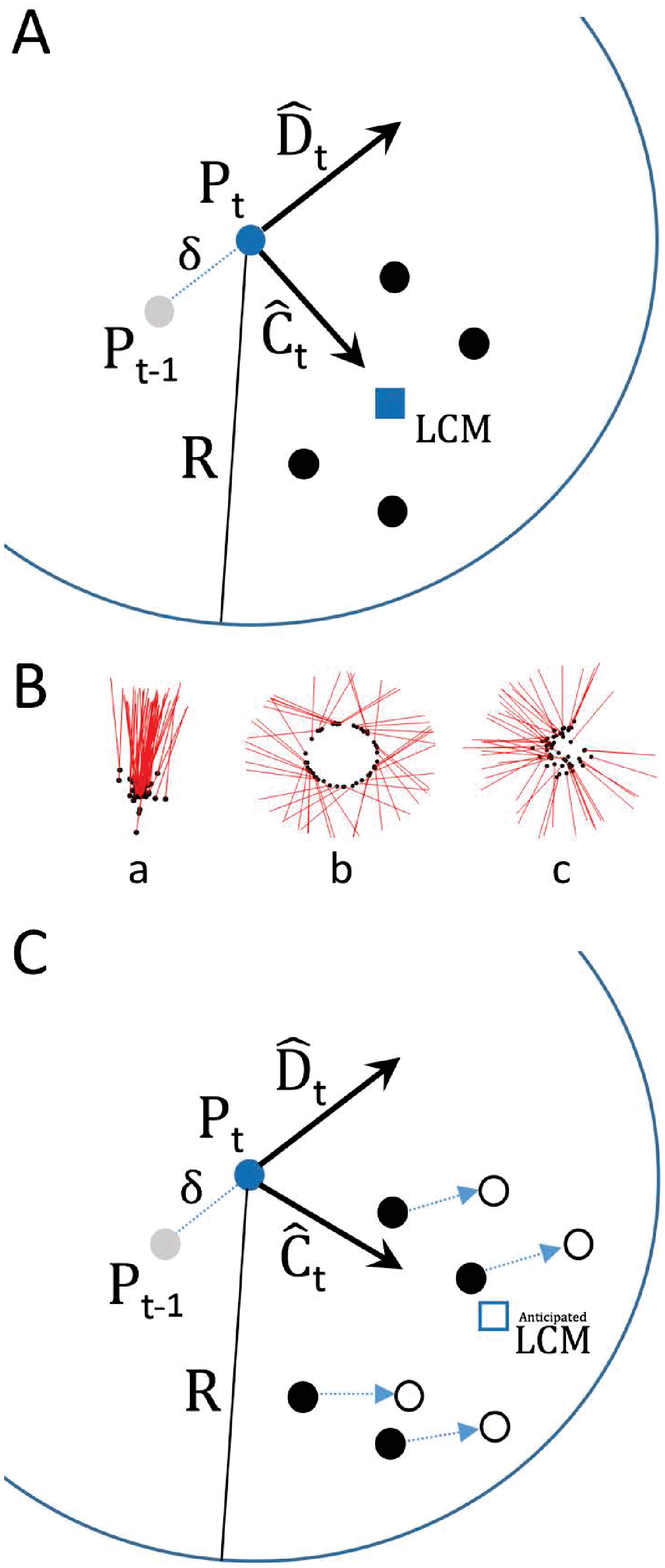}
\caption{Illustration of how a focal particle calculates its new heading on each time step in the synchronous LAM. (A) Model without anticipation. (C) Model with anticipation. Dots represent the current positions of nearby particles and the rings their anticipated future positions. (B) Group types observed in the models (a) cohesive polarized group, (b) mill, (c) swarm.}
\label{fig:1}
\end{figure}

We use the simplest form of the synchronous LAM \cite{Strombom2017}. This is a self-propelled particle (SPP) model in which particles interact via local attraction alone and update their headings and positions synchronously. More specifically, on every time step $t$ each particle calculates the position of the local center of mass (LCM) of other particles within a distance of $R$ from it. Its new heading ($\bar{D}_{t+1}$) is given by a linear combination of the normalized direction toward the LCM ($\hat{C}_t$) and its current normalized heading ($\hat{D}_t$). 
\begin{equation} \bar{D}_{t+1}=c\hat{C}_t + \hat{D}_t.\end{equation}
See Fig. \ref{fig:1}A. The main model parameter $c$ specifies the relative tendency to steer towards the LCM when the relative tendency to proceed with the current heading is $1$. Once a new heading has been calculated for each particle all particles are simultaneously moved a distance of $\delta$ in their respective headings. Depending on the relative strength of local attraction ($c$) groups of different type will be produced by the model. In the synchronous LAM we see the following regimes. If $c<0.2$ no group will form. If $0.2<c<1.7$ mills will form (Fig. \ref{fig:1}Bb). If $c>1.7$ swarms will form (Fig. \ref{fig:1}Bc). When anticipation is included each particle calculates the LCM from the anticipated positions of particles nearby. See Fig. \ref{fig:1}C. The anticipated position of a nearby particle is the position that the particle would be at if it continued with its current heading for a distance of $\delta$. If the anticipated future position of a nearby particle is within a distance of $R$ from the focal particle it counts as a neighbor and will influence the calculation of the LCM.

Our goal is to investigate the causal effects of including anticipation in the synchronous LAM described and analyzed in \cite{Strombom2017}. Therefore we perform the same simulations and analysis of the model with anticipation here. In summary, we ran 100 simulations for each $c$ from 0.04 to 2 in increments of 0.02 in the synchronous LAM with anticipation and kept all other parameters fixed at $N=50$, $R=4$, $\delta=0.5$. Periodic boundary conditions were used and each particle was assigned a random position and heading at the start of each simulation. We measured the polarization ($\alpha$) and scaled size ($\sigma$) of the resulting group. The polarization measures the degree to which the $N$ particles are moving in the same direction and ranges from 0 to 1 \cite{Vicsek1995}. Scaled area provides a measure of how much of the available space the group occupies and ranges from 0 to 1 \cite{Strombom2011}. Combining these two measures allows us to distinguish between the three groups in Fig 1B and the case when no group formed. If no group formed $\sigma$ is high and $\alpha$ is intermediate, cohesive polarized groups have low $\sigma$  and high $\alpha$ ($\approx 1$), mills have a low $\sigma$ value that decreases with $c$ and very low $\alpha$, and swarms have a very low $\sigma$ and low-intermediate $\alpha$. See the Model and Simulations and the Materials and Methods sections of \cite{Strombom2017} for more detailed information about the model, simulations and measures.

\section*{Results} 
\begin{figure}[h!]
\centering
\includegraphics[width=\columnwidth]{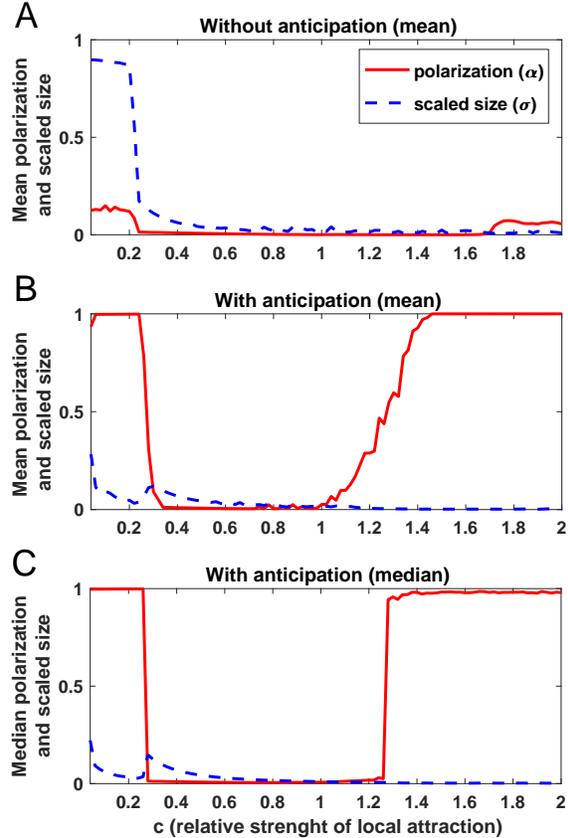}
\caption{Mean/Median polarization ($\alpha$) and scaled size ($\sigma$) over 100 simulations with $c$ from 0.04 to 2. (A) Model without anticipation (From \cite{Strombom2017}). (B) Model with anticipation.}
\label{fig:2}
\end{figure}

Including anticipation affects the model behavior dramatically in some parameter regions and leaves it effectively unchanged in others. For $c$ less than 0.2 the model without anticipation generates no group ($\sigma$ high, $\alpha$ intermediate) whereas the model with anticipation generates cohesive polarized groups ($\sigma$ low, $\alpha$ high). For $c$ between 0.2 and 1 both models generate mills ($\sigma$ low and decreasing with $c$, $\alpha$ very low). For $c$ larger than 1 the models generate different results. The model without anticipation reliably produces mills up to about $c=1.4$ and then swarms ($\sigma$ low, $\alpha$ low-intermediate) whereas the model with anticipation starts producing cohesive polarized groups again in this regime. The correct interpretation of the gradual increase in polarization $\alpha$ for $c$ between 1 and 1.4 seen in Fig. \ref{fig:2}B is that the proportion of simulations that produce cohesive polarized groups instead of mills increases with $c$. Fig. \ref{fig:2}C shows the median of the polarization and scaled size measures and we see that there is a sharp transition in the polarization measure at around $c=1.25$  consistent with this claim.

\section*{Discussion}
We have established that including anticipation in a synchronous version of the LAM \cite{Strombom2017} induces cohesive polarized group formation, inhibits swarm formation and leaves part of the mill regime unaffected while destabilizing other parts (Fig. \ref{fig:2}). This in contrast to the findings that including anticipation in alignment based models, and one attraction-repulsion model, promotes milling and swarming and inhibits polarized collective motion \cite{Morin2015,Gerlee2016,Baggaley2016}.

The finding that anticipation induces polarized collective motion in several disjoint parts of the parameter space is particularly surprising. Not only does anticipation destabilize swarms and smaller mills in favor of cohesive polarized groups. It also enables cohesive polarized groups to reliably form where no group would form in the absence of anticipation. Hence, at least in models similar to ours where collective motion is driven by local attraction anticipation may be an underutilized mechanism for inducing polarized collective motion. This observation may also be useful to consider in the context of inferring interaction rules in real animal groups. In particular, where some studies find no evidence for alignment forces acting in schools of fish \cite{Teddy2011,Katz2011} while others do \cite{Gautrais2012}. Of course there could be many other circumstances that explain this, as discussed in \cite{Strombom2015}. However, adding anticipation to the list of mechanisms that can generate polarized collective motion in SPP models, as a complement to asynchrony \cite{Strombom2011,Strombom2017}, asymmetric interactions, explicit alignment terms, and more, may be useful in advancing our understanding of how qualitatively similar group level collective motion emerges from seemingly different local interactions between individuals in the group.

As our model is a synchronous, off lattice model that employs neighborhood-wide anticipation we now know that asynchrony, motion restricted to a lattice, and mutual anticipation are not necessary requirements for anticipation to induce or promote polarized collective motion. Combining our results with the findings in \cite{Murakami2017} suggests that the inducing effect of anticipation polarized collective motion is a more general phenomenon than previously thought because a similar effect is observed under a variety of different assumptions. Furthermore, factoring in the results from \cite{Morin2015,Gerlee2016,Baggaley2016}, it appears that the effects of including anticipation in a model is itself hard to anticipate. Because it can have a similar effect on different models, e.g. alignment based and attraction-repulsion models, but also different effects on similar models, e.g. attraction only and attraction-repulsion models. This suggests that it might be useful to implement anticipation in other models of collective motion. In particular the full attraction, repulsion and orientation model \cite{Couzin2002} and other attraction-repulsion models \cite{Romanczuk2009,Romanczuk2012,Strombom2015}. Perhaps including anticipation in these would enable them produce previously unobserved group types and phenomena. For example, the transition behavior and multi-stability observed in \cite{Tunstrom2013}.

In addition to the potential usefulness of considering anticipation in the context of collective motion in groups of homogeneous individuals it may also be beneficial in the context of leader-follower and herding systems. From \cite{Strombom2017} we know that even partial asynchrony induces polarized collective motion and it was speculated that perhaps leaders, or informed, individuals update more asynchronously and followers update more synchronously. Similarly, it seems reasonable to suspect that, if employed at all, leaders would rely on anticipation less and followers rely on anticipation more. Conversely, in herding situation we may expect the shepherd to rely heavily on anticipation. Both of these predictions might be testable is suitable cases where high resolution tracking data is available and theoretically by implementing anticipation in the model of leadership in \cite{Couzin2005} and the shepherding model in \cite{Strombom2014} and other similar models. Perhaps attraction and anticipation and/or asynchrony are the main driving forces operating in some real world leader-follower or herding systems.

\section*{Acknowledgments}
This work was supported by a grant from the Swedish Research Council to D.S. (ref: 2015-06335). 

\bibliographystyle{vancouver}
\bibliography{bibantic}

\end{document}